\documentclass[aps,pra,groupedaddress,amsfonts,amssymb]{revtex4}
\usepackage{amsmath}
\usepackage{amsfonts}
\usepackage{amssymb}
\usepackage{epsfig}
\newtheorem{The}{Theorem}

\newtheorem{Pro}[The]{Proposition}

\newenvironment{proof}{\par\noindent\textit{Proof.\ }}{\hfill $\square$ \vspace{1em}}

\def\idty{\openone}

\def\ketbra#1#2{\vert#1\rangle\langle#2\vert}

\newcommand{\tr}{\operatorname{tr}}
\def\id{{\rm id}}

\def\HH{{\cal H}}\def\BB{{\cal L}}
\newcommand{\e}{\mathrm{e}}
\newcommand{\im}{\mathrm{i}}

\newcommand{\prob}{\mathbb{P}}

\def\Tcorr{T_{\rm corr}}
\def\Scorr{S_{\rm corr}}

\newcommand{\fid}{\mathcal{F}}

\begin{document}
%* Top Matter
\title{On quantum error-correction by classical feedback in discrete time}
\author{M. Gregoratti}
\email{gregoratti@mate.polimi.it}
\affiliation{Dip.\ Mat., Politecnico di Milano, piazza Leonardo da Vinci 32, I-20133 Milano, Italy}
\author{R.~F. Werner}
\email{R.Werner@TU-BS.DE}
\affiliation{Inst.\ Math.\ Phys., TU-Braunschweig, Mendelssohnstra{\ss}e 3, D-38106 Braunschweig,
Germany}
\begin{abstract}
We consider the problem of correcting the errors incurred from sending
quantum information through a noisy quantum environment by using classical
information obtained from a measurement on the environment.
For discrete time Markovian evolutions,
in the case of fixed measurement on the environment,
we give criteria for quantum information to be perfectly corrigible and
characterize the related feedback. Then we analyze the case when perfect correction is not possible
and, in the qubit case, we find optimal feedback maximizing the channel fidelity.
\end{abstract}
\maketitle
%%%%%%%%%%%%%%%%%%%%%%%%%%%%%%%%%%%%%%%%%%%%%%%%%%%%%%%%%%%%%%%%%%%%%%%%%%%%%%%%%%%%%%%%%%%%%%%%%
%\special{!userdict begin /start-hook{gsave 180 80
%translate 65 rotate /Times-Roman findfont 80 scalefont setfont 50
%0 moveto 0.8 setgray (Version 10.07.2002) show grestore}def end}
%*%%%%%%%%%%%%%%%%%%%%%%%%%%%%%%%%%%%%%%%%%%%%%%%%%%%%%%%%%%%%%%%%%%%%%%%%%%%%%%%%%%%%%%%%%%%%%%%%%%
\section{Introduction}

Error correction is a key problem in quantum information processing. Without it, decoherence would easily
destroy all hopes for quantum computation and quantum cryptography.

Two important approaches have been devoloped to combat decoherence. The usual theory of quantum
error-correction \cite{NC,KW} redundantly encodes the original quantum information in a larger quantum
system by a unitary operation
which maps the initial Hilbert space into the code space, a subspace of the Hilbert space
associated to this larger system. After encoding, the larger system is subjected to noise and then a
measurement is performed on the system to diagnose the type of error which occurred. Finally, on the
basis of the outcome, a restoring operation is performed to return the system to the original state in
the code space.

Another approach \cite{MZ,Plenio,ADL,AWM,Alber}, based on feedback control, has been developed for
systems which are continuously monitored during
their noisy evolution. It employs the result of this continuous measurement to determine the errors
occurred and, on this basis, perform corrections in real time to protect
states which are known to lie initially within a certain code space.
The continuous measurement can be performed introducing an additional interaction to the evolution of
the system or preferably by simply observing the environment after its interaction with the system.
Up to now there is no general theory for quantum feedback control, but some correction schemes
have been considered and have dealt only with Hamiltonian feedback.

Taking from feedback control the idea of a correction scheme based on a measurement performed not on
the system but on the environment, we have analyzed in an earlier publication \cite{GW} the
different possible behaviours of a channel with respect to the
existence of a measurement allowing perfect correction of
information, quantum or classical. No code space is introduced and, for every initial state, the aim is
to find proper measurement and restoring operations, not necessarily unitary, to recover the initial
information or at least optimally restore it.

Here we develop the analysis in \cite{GW} to evolutions composed of many time steps.
We consider quantum information carried by a system undergoing a discrete time Markovian evolution
(multi-step channel $T=T^{(n)}\circ\cdots\circ T^{(1)}$) and allow correction operations between any
two subsequent steps: every time we perform a measurement on the environment and
a restoring operation on the system. We take the measurments on the environment as given
and we look for the best restoring operations choosen on the basis of all the outcomes observed so
far. First we give conditions for quantum information to be perfectly corrigible and
characterize the related feedback.
It turns out that classical feedback is not useful because perfect correction of quantum information
is possible only if a unique correction at the end would suffice.
Then we analyze the case when perfect correction is not possible
and, in the qubit case, we find optimal feedback maximizing the channel fidelity.
Similarly to the case of perfect correction, but more surprisingly,
classical feedback is still useless if quantum information carried by one qubit
because again a unique restoring operation performed at the end of the evolution guarantees the same
performance of correction. Anyway
classical feedback is shown to be helpful when the quantum carrier has a larger Hilbert
space.

Every Section of the paper considers first the case of quantum information sent through a single-step
channel, so that it is possible only one correction after this noisy evolution just as in \cite{GW},
and then the case of a multi-step
channel, so that classical feedback is possible during the noisy evolution.
In Section~\ref{CorrSch}, we set up the framework and the
basic correction scheme. In Section~\ref{PerCorr} we prove the basic criteria for the
existence of a correction scheme for a given channel and given
measurements on the environment, and we characterize the related feedback. In Section~\ref{OptRec}
we look for optimal correction scheme when it is not be possible to achieve a
complete correction of errors and ...

%*%%%%%%%%%%%%%%%%%%%%%%%%%%%%%%%%%%%%%%%%%%%%%%%%%%%%%%%%%%%%%%%%%%%%%%%%%%%%%%%%%%%%%%%%%%%%%%%%%%

\section{The correction scheme}\label{CorrSch}

We consider quantum information carried by a quantum system with finite dimensional Hilbert space $\HH$.
We work in the Schr\"odinger picture, so the action of the noisy \textit{channel},
which corrupts the information transforming each
input density operator $\rho$ on $\HH$ to a different output density operator $T(\rho)$ on $\HH$,
is given by a
completly positive, trace preserving, linear map $T:\BB(\HH)\to\BB(\HH)$, where $\BB(\HH)$ denotes the
space of all linear operators on $\HH$.
A channel $T$ is not physically reversible, i.e.\ there is no channel $R$ such that $R\circ T=\id$, unless
$T$ is a unitary channel $T(\rho)=u\,\rho\,u^*$, with $u\in\mathcal{U}(\HH)$, the group of unitary
operators on $\HH$.

\subsection{Single-step channel}

Every channel can be described as the result of a unitary coupling to an environment, followed by the
discard of the environment after the interaction. Nevertheless, the initial information carried by
$\HH$ is then shared with the environment and indeed, if we controlled perfectly the combined system,
then we could recover the initial quantum information simply by reversing the
global evolution, so restoring perfectly the input state $\rho$.
We do not hypothesize this, but we assume that our control on the environment is good enough to have its
initial state pure and to be able to perform any measurement on it after the interaction with $\HH$.
The measurement of an observable $X$ on the environment decomposes the channel $T$ into an
\textit{instrument}, a
family of completely positive maps $T_x$ giving the (non-normalized) output
states $T_x(\rho)$ of the subensembles of systems selected according to the result ``$x$'' of
this measurement:
the probability of observing $x$ is $\prob(X=x)=\tr T_x(\rho)$, the normalized output state for the
corresponding subensemble is $T_x(\rho)/\tr T_x(\rho)$, and the
expectation in that subensemble of a self-adjoint $A\in\BB(\HH)$ is $\tr(T_x(\rho) A)/\tr T_x(\rho)$.
Ignoring the result of the measurement one recovers the original channel
\begin{equation}\label{T&Tx}
    T=\sum_x T_x\;.
\end{equation}
Of course, the result $x$ of the measurement gives classical information about the environment after
the interaction, and hence also about $\HH$ before the interaction. The idea for the correction scheme is
to employ this information to select a proper restoring channel: we introduce a family of channels
$R^{(x)}:\BB(\HH)\to\BB(\HH)$ where $R^{(x)}$ can depend in an arbitrary way on $x$.
After correction, the state of
the subensemble for which  the measurement has given the  result $x$, will be
$R^{(x)}(T_x(\rho))$ up to the normalization factor $\tr(T_x(\rho))$. The overall
corrected channel is built from these conditional operations by ignoring the intermediate
information $x$, and is the sum of these contributions:
\begin{equation}\label{Tcorr}
  \Tcorr=\sum_x  R^{(x)}\circ T_x\;.
\end{equation}

The goodness of the scheme in restoring quantum information depends on how $\Tcorr$ can be brought
close to the ideal channel on $\HH$, i.e. $\id$.

Whether or not we can find a good correction scheme in principle depends not only on the noisy channel
$T$, but also on the set of decompositions \eqref{T&Tx} obtainable by a measurment on the environment,
which usually depends on the particular coupling which induces the noisy evolution $T$. Anyway the
assumption of a pure environment overcomes this problem because it
guarantees that every decomposition of $T$ into
c.p.\ summands $T_x$ can be realized by a measurement on the environment.
Moreover, in order to correct quantum information, the
preferable docompositions are the finest ones, those for which no proper refinement is possible.
Therefore we shall consider the non refinable decompositions given by the Kraus representations of $T$:
\begin{equation}\label{Tdec}
  T(\rho) = \sum_x t_x \, \rho \, t_x^* \;, \qquad \mbox{ where } \sum_x t_x^*t_x = \idty \;.
\end{equation}
In the following we shall assume that a decomposition \eqref{Tdec} is given, that it is associated to the
measurement of an observable $X$, and for it we shall consider
the problem of perfectly correcting quantum information or, if this is not possible, optimally restoring
it.

\subsection{Multi-step channel}

We are especially interested in multi-step channels, where the noisy evolution $T$ is given by
$n$ Markovian steps:
\begin{equation}\label{divT}
    T=T^{(n)}\circ\cdots\circ T^{(1)}\;,
\end{equation}
where $T^{(k)}:\BB(\HH)\to\BB(\HH)$ are $n$ possibly different channels.
In this case we assume to be able to perform a
measurement at each step, so that every $T^{(k)}$ is decomposed into
\begin{equation}\label{Tkdec}
    T^{(k)}(\rho) = \sum_{x_k} t^{(k)}_{x_k} \, \rho \, {t^{(k)}_{x_k}}^* \;,
    \qquad \mbox{ where } \sum_{x_k} {t^{(k)}_{x_k}}^*t^{(k)}_{x_k} = \idty \;,
\end{equation}
the channel $T$ is decomposed into
\begin{equation}\label{divTdec}
    T(\rho) = \sum_{x_1,\ldots,x_n} t^{(n)}_{x_n} \cdots t^{(1)}_{x_1}\, \rho \, {t^{(1)}_{x_1}}^*
    \cdots {t^{(n)}_{x_n}}^*\;,
\end{equation}
and the probability of observing $(x_1,\ldots,x_n)$ is $\prob(X_1=x_1,\ldots,X_n=x_n) =
\tr t^{(n)}_{x_n} \cdots t^{(1)}_{x_1}\, \rho \, {t^{(1)}_{x_1}}^* \cdots {t^{(n)}_{x_n}}^*$.

\begin{figure}[htb]
\epsfxsize=12cm \epsffile{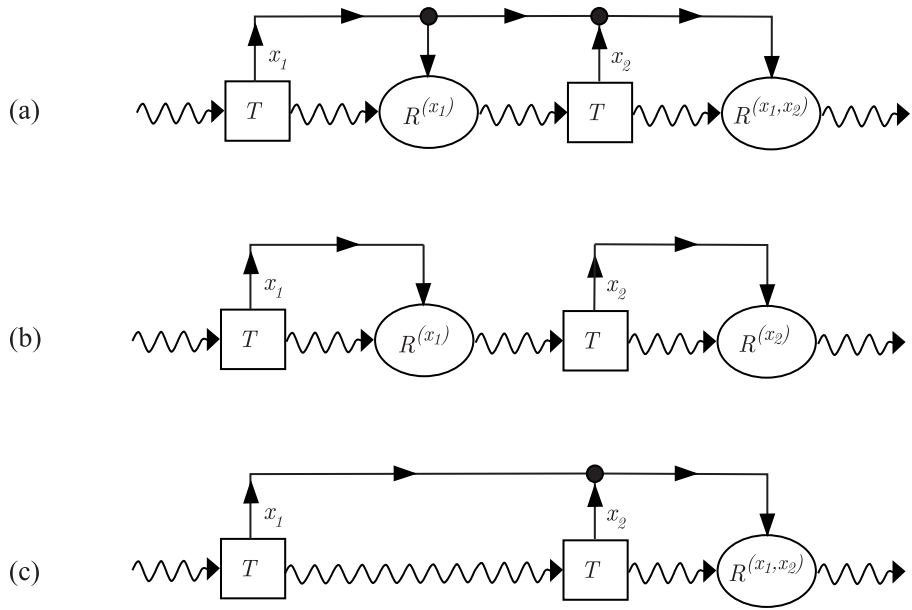}
 \caption{\label{schemes}{\it Different error correction schemes for multiple channels.
  (a) represents the most general case. In (b) only the information from the immediately
  preceding step is used. In (c) only one correction is done at the end.
  %, depending, however on all channel outputs.
  }}
\end{figure}

We also assume to be able to interfere in the evolution of the system after each step by appling a
restoring channel selected according to the whole information gathered so far.
For every $k=1,\ldots,n$, we introduce a family of channels
$R^{(x_1,\ldots,x_k)}:\BB(\HH)\to\BB(\HH)$, where $R^{(x_1,\ldots,x_k)}$ can depend in an arbitrary
way on $(x_1,\ldots,x_k)$ and it is applied after the measurement of $X_k$ so that the overall corrected
channel turns out to be
\begin{equation}\label{divTcorr}
  \Tcorr(\rho)= \sum_{x_1,\ldots,x_n} R^{(x_1,\ldots,x_n)}\Big(t^{(n)}_{x_n} \cdots R^{(x_1)} \big(
  t^{(1)}_{x_1}\, \rho \, {t^{(1)}_{x_1}}^* \big) \cdots {t^{(n)}_{x_n}}^* \Big) \;.
\end{equation}
So the classical information obtained by the measurement process
on the environment is fedback to modify the evolution of the
system $\HH$. Notice that this general scheme includes the special
cases (compare Fig.~\ref{schemes}):
\begin{itemize}
\item $R^{(x_1,\ldots,x_k)}= R^{(x_k)}$ for every $k=1,\ldots,n$,
i.e.\ every restoring channel is selected according only to the
last observation $x_k$.
 \item $R^{(x_1,\ldots,x_k)}=\id$ for every
$k=1,\ldots,n-1$, i.e.\ the information given by the measurement
on the environment is gathered during the evolution, but every
correction is deferred after it,
\end{itemize}

In the following we shall assume that a decomposition \eqref{divTdec} is fixed, that it is associated to
a measurement process, and, given this, we shall
consider the problem of perfect or optimal correction of quantum information.

\section{Perfect correction of Quantum Information}\label{PerCorr}

Given a channel $T$ with a fixed decomposition \eqref{Tdec} or \eqref{divTdec}, our scheme can
perfectly correct quantum information if one can find restoring channels $R$ such that
\begin{equation}\nonumber
    \Tcorr=\id\;.
\end{equation}

\subsection{Single-step channel}

Let us improve the criterion given in \cite{GW}.
\begin{Pro}\label{PCQI} Let $T:\BB(\HH)\to\BB(\HH)$ be a channel over a finite dimensional Hilbert
space $\HH$. Given a Kraus decomposition $T(\rho)=\sum_x t_x\,\rho\,t_x^*$, the following are equivalent
\begin{itemize}
\item[(a)] there exists a family of channels $R^{(x)}:\BB(\HH)\to\BB(\HH)$ perfectly restoring quantum
information;
\item[(b)] $t_x^*t_x=p_x\,\idty$ for all $x$, with $p_x\geq0$, $\sum_x p_x=1$;
\item[(b')] $t_x=\sqrt{p_x}\,u_x$ for all $x$, with $u_x\in\mathcal{U}(\HH)$, $p_x\geq0$, $\sum_x p_x=1$;
\item[(b'')] $T(\rho) = \sum_x p_x \,u_x\,\rho\,u_x^*$ is a convex combination of unitary channels;
\item[(c)] $\prob(X=x)=\tr t_x\,\rho\,t_x^*$ is independent of the input state $\rho$.
\end{itemize}
When these conditions hold, then the restoring channels in (a) have to be
\begin{equation}\label{PF}
    R^{(x)}(\rho)=u_x^*\,\rho\,u_x \;,
\end{equation}
and the probability law of the outcome is
\begin{equation}\label{opl}
    \prob(X=x) = p_x \;.
\end{equation}
\end{Pro}
\begin{proof} (a) $\Rightarrow$ (b). Let $R^{(x)}$ be channels such that $\sum_x  R^{(x)}
(t_x\,\rho\,t_x^*) = \rho$ for every state $\rho$. Then for every $x$, by the ``No information without
perturbation'' Theorem, $R^{(x)}(t_x\,\rho\,t_x^*) = p_x \, \rho$, for some $p_x\geq0$,
$\sum_x p_x=1$. Therefore $\tr t_x^*t_x\,\rho = \tr t_x\,\rho\,t_x^* = \tr R^{(x)}(\tr
t_x\,\rho\,t_x^*) = p_x$ for all $\rho$, and so $t_x^*t_x=p_x\,\idty$.

\medskip

(b) $\Leftrightarrow$ (b') $\Leftrightarrow$ (b''). These conditions are clearly all equivalent because
$\dim \HH < \infty$.

\medskip

(b') $\Rightarrow$ (c). $\prob(X=x) = \tr t_x\,\rho\,t_x^* = \tr
\sqrt{p_x}\,u_x^*\,\rho\,\sqrt{p_x}\,u_x = p_x$ for every $\rho$, so (b') implies (c), and equation
\eqref{opl}, too.

\medskip

(c) $\Rightarrow$ (a). $\tr t_x^*t_x\,\rho = \tr t_x\,\rho\,t_x^* = \prob(X=x)$ which does not depend on
$\rho$ and so $t_x^*t_x=\prob(X=x)\,\idty$. Then $t_x = \sqrt{\prob(X=x)}\,u_x$, where $u_x$ is
unitary because $\dim \HH<\infty$, and the channels $R^{(x)}(\rho)=u_x^*\,\rho\,u_x$
perfectly restore quantum information.

\medskip

Now we have to show that conditions (a), (b) and (c) imply \eqref{PF}.
Consider a family of channels $R^{(x)}$ restoring quantum information, with
$R^{(x)}(\rho)=\sum_j r^{(x)}_j\,\rho\,{r^{(x)}_j}^*$ and $\sum_j {r^{(x)}_j}^*r^{(x)}_j= \idty$. Then
\begin{equation}\nonumber
    \sum_{x,j} r^{(x)}_j\,t_x\,\rho\,t_x^*\,{r^{(x)}_j}^* = \rho\;,\qquad\forall\;\rho\;,
\end{equation}
and using again the ``No information without perturbation'' Theorem we have
\begin{equation}\nonumber
    r^{(x)}_j\,t_x = \sqrt{p_{x,j}}\,\e^{\im\theta(x,j)}\,\idty\;, \qquad \forall\;x,j,
\end{equation}
for some probability law $p_{x,j}$ and some real function $\theta$. Using (b'), we get
\begin{equation}\nonumber
    r^{(x)}_j = \sqrt{\frac{p_{x,j}}{p_x}}\,\e^{\im\theta(x,j)}\,u_x^* \propto u_x^*
    \qquad \forall\;j,
\end{equation}
i.e., fixed $R^{(x)}$, all the Kraus oprators $r^{(x)}_j$ have to be proportional to the same
unitary operator $u_x^*$, and \eqref{PF} holds.
\end{proof}

Let us remark that condition (b) gives a simple criterion for the existence of channels $R$ perfectly
restoring quantum information based directly on the Kraus operators of $T$.
Moreover it clarifies the structure of channels which allow perfect correction with our scheme:
the measurement on the environment has to decompose $T$ into a convex
combination of unitary channels, so that the overall evolution can be seen as an average of reversible
evolutions $\rho\mapsto u_x\,\rho\,u_x^*$, which occur randomly with probability
$p_x$; the measurement detects the transformation occurred and $R^{(x)}(\rho') =
u_x^*\,\rho'\,u_x$ restores the initial state $\rho$. This means also, by contition (c),
that the measurement on the environment is completely uninformative about the system $\HH$:
the output $x$ can be interpreted as
classical information about "what happened" to the system $\HH$, but it gives no information about the
input system because it is independent of its state. And the uninformative character of the observation,
if associated to a Kraus decomposition of $T$, is sufficient for perfect correction.
Moreover, if perfect correction is possible, the restoring channels $R$ have to be untiary.

We shall call \textit{corrigible} every channel $T$ with a decomposition satisfying condition (b).

Just to show how useful can be a suitable measurement on the environment,
let us remark that our scheme can perfectly correct quantum
information sent through a depolarizing channel, which is known to destroy all quantum
information and for which no ordinary quantum error correcting code works (it has zero quantum and
classical capacity). Indeed, set $N = \dim\HH$, the depolarising
channel $T(\rho)= \frac{1}{N} \idty$ admits the Kraus representation
\begin{equation}\nonumber
    T(\rho) = \sum_{x,y=1}^N t_{x,y} \,\rho\, t_{x,y}^*\;, \qquad
    t_{x,y} = \frac{1}{N} \sum_{z=1}^N \e^{\frac{2\pi\im}{N}zy} \, \ketbra{z+x}{z}
            = \frac{1}{N}\,  u_{x,y}\;,
\end{equation}
with unitary operators $u_{x,y}$, where  $\{|z\rangle\}_{z\in\mathbb{Z}_N}$ denotes a  basis
labeled cyclically so that addition in $|z+x\rangle$ is modulo $N$.

\subsection{Multi-step channel}

Let us consider now perfect correction for multi-step channels.
\begin{Pro}\label{divPCQI} Let $T=T^{(n)}\circ\cdots\circ T^{(1)}$,
with every $T^{(k)}:\BB(\HH)\to\BB(\HH)$, be a multi-step channel over a finite dimensional Hilbert
space $\HH$. Given the Kraus decompositions
$T^{(k)}(\rho)=\sum_{x_k} t^{(k)}_{x_k}\,\rho\,{t^{(k)}_{x_k}}^*$, T.F.A.E.\
\begin{itemize}
\item[(a)] there exists a family of channels $R^{(x_1)}$, $R^{(x_1,x_2)}$, \ldots, $R^{(x_1, \ldots,
x_n)}$ perfectly restoring quantum information;
\item[(b)] every channel $T^{(k)}(\rho)=\sum_{x_k} t^{(k)}_{x_k}\,\rho\,{t^{(k)}_{x_k}}^*$ is corrigible,
\\
\noindent i.e.\ for all $x_k$ we have $t^{(k)}_{x_k}= \sqrt{p^{(k)}_{x_k}}\,u^{(k)}_{x_k}$, with
$u^{(k)}_{x_k}\in\mathcal{U}(\HH)$, $p^{(k)}_{x_k}\geq0$, $\sum_{x_k} p^{(k)}_{x_k}=1$;
\item[(c)] $\prob(X_1=x_1,\ldots,X_n=x_n) = \tr t^{(n)}_{x_n} \cdots t^{(1)}_{x_1}\, \rho \,
{t^{(1)}_{x_1}}^* \cdots {t^{(n)}_{x_n}}^*$ is independent of the input state $\rho$.
\end{itemize}
When these conditions hold, then for every $k=1, \ldots, n$ the restoring channels in (a) have to be
\begin{equation}\label{divpF}
    R^{(x_1, \ldots, x_k)}(\rho)=v^{(x_1, \ldots, x_k)}\,\rho\,{v^{(x_1, \ldots, x_k)}}^* \;, \qquad
    v^{(x_1, \ldots, x_k)} \in \mathcal{U}(\HH)\;,
\end{equation}
with
\begin{equation}\label{divLpF}
    v^{(x_1, \ldots, x_n)}
    = {u^{(1)}_{x_1}}^*\,{v^{(x_1)}}^*\cdots {v^{(x_1, \ldots, x_{n-1})}}^*\,{u^{(n)}_{x_n}}^*\;,
\end{equation}
and the probability law of the outcome process is always
\begin{equation}\label{divopl}
    \tr R^{(x_1,\ldots,x_n)}\Big(t^{(n)}_{x_n} \cdots R^{(x_1)} \big( t^{(1)}_{x_1}\, \rho \,
    {t^{(1)}_{x_1}}^* \big) \cdots {t^{(n)}_{x_n}}^* \Big) = \prod_{k=1}^n p^{(k)}_{x_k}
\end{equation}
\end{Pro}

Therefore, as long as we are interested in perfect correction of quantum information, the multi-step
structure of the channel and the possibility of applying feedback during the evolution do not help:
perfect correction is possible if and only if every step is corrigible and, in this case, it is enough to
make a unique correction at the end. Indeed only unitary corrections are allowed, otherwise the original
information would be corrupted by the feedback itself, and then the first $n-1$ unitaries
$v^{(x_1, \ldots, x_k)}$ can be chosen arbitrarly, also $v^{(x_1, \ldots, x_k)}=\idty$, provided that
the whole evolution is reversed by the last one (eq. \eqref{divLpF}). Again, the channel is corrigible
if and only if the measured process is uninformative; moreover its probabilistic law,
which cannot be modified by the unitary feedback, is that of a sequence of independent random variables.

The proof of Proposition \ref{divPCQI} goes via a more general result about the
composition of $n$ channels when each channel is decomposed according to a measurement and, not only each
channel itself, but also its decomposition, may depend on the previous observations. Therefore, denoted
by $x_k$ the outcomes of the measurement at the $k^\mathrm{th}$ step, we want to consider the evolution
of a quantum system sent through a sequence of channels on $\BB(\HH)$
\begin{equation}\label{InstrProc}
    T^{(x_1,\ldots,x_{k-1})}(\rho)=
    \sum_{x_k} t^{(x_1,\ldots,x_{k-1})}_{x_k}\,\rho\,{t^{(x_1,\ldots,x_{k-1})}_{x_k}}^*\;, \qquad
    \sum_{x_k}{t^{(x_1,\ldots,x_{k-1})}_{x_k}}^*t^{(x_1,\ldots,x_{k-1})}_{x_k} = \idty\;,
\end{equation}
where for the first step, $k=1$, we denote by ``0'' the empty string of prior results, so we write
\begin{equation}
    T^{(0)}(\rho)=\sum_{x_1} t^{(0)}_{x_1}\,\rho\,{t^{(0)}_{x_1}}^*\;, \qquad
    \sum_{x_1}{t^{(0)}_{x_1}}^*t^{(0)}_{x_1} = \idty \;.\nonumber
\end{equation}
Then the total evolution is
\begin{equation}\label{divTdec0}
    T(\rho) = \sum_{x_1,\ldots,x_n} t^{(x_1,\ldots,x_{n-1})}_{x_n} \cdots t^{(0)}_{x_1}\,
    \rho \, {t^{(0)}_{x_1}}^* \cdots {t^{(x_1,\ldots,x_{n-1})}_{x_n}}^*\;.
\end{equation}
and the next Proposition holds.
\begin{Pro}\label{divQIrest0} Given a finite dimensional Hilbert
space $\HH$, let $T:\BB(\HH)\to\BB(\HH)$ be a channel \eqref{divTdec0} resulting from the application
of $n$ decomposed channels $T^{(x_1,\ldots,x_{k-1})}$ \eqref{InstrProc}. If $T = \id$, then
\begin{itemize}
\item[(a)] every channel $T^{(x_1,\ldots,x_{k-1})}(\rho)= \sum_{x_k}
t^{(x_1,\ldots,x_{k-1})}_{x_k}\,\rho\,{t^{(x_1,\ldots,x_{k-1})}_{x_k}}^*$ is corrigible;
\item[(b)] a channel $T^{(x_1,\ldots,x_{k-1})}$ is unitary if later channels and their decomositions
do not depend on the value $x_k$ observed at the $k^{\mathrm{th}}$ step.
%(automatically true for $T^{(x_1,\ldots,x_{n-1})}$).
\end{itemize}
\end{Pro}
\begin{proof}
If $T = \id$, then
\begin{equation}\label{f1}
    t^{(x_1,\ldots,x_{n-1})}_{x_n} \cdots t^{(0)}_{x_1} = \sqrt{p(x_1,\ldots,x_n)} \,
    \e^{\im\theta(x_1,\ldots,x_n)} \, \idty \;, \qquad \forall\;x_1,\ldots,x_n,
\end{equation}
where $p$ is the joint probability law of the outcomes, independent of $\rho$, and $\theta$ is a real
function.

(a) Using the normalization property of Kraus operators, we can immediately show that $T^{(0)}$ is
corrigible:
\begin{equation}\nonumber
    {t^{(0)}_{x_1}}^*t^{(0)}_{x_1} =
    \sum_{x_2,\ldots,x_n} {t^{(0)}_{x_1}}^*\,{t^{(x_1)}_{x_2}}^* \cdots
    {t^{(x_1,\ldots,x_{n-1})}_{x_n}}^*t^{(x_1,\ldots,x_{n-1})}_{x_n} \cdots t^{(x_1)}_{x_2} \,
    t^{(0)}_{x_1} = \sum_{x_2,\ldots,x_n} p(x_1,\ldots,x_n) \, \idty = p(x_1) \, \idty\;,
\end{equation}
where $p(x_1)$ is the probability of observing $x_1$ at the first step. Now, since
$t^{(0)}_{x_1} = \sqrt{p(x_1)}\,u^{(0)}_{x_1}$ for some unitary $u^{(0)}_{x_1}$, we can fix $x_1$ and
find that $T^{(x_1)}$ is corrigible:
\begin{equation}\nonumber\begin{split}
    {t^{(x_1)}_{x_2}}^*t^{(x_1)}_{x_2} &= \frac {1}{p(x_1)}\, u^{(0)}_{x_1} \,
    {t^{(0)}_{x_1}}^*\,{t^{(x_1)}_{x_2}}^*t^{(x_1)}_{x_2}\,t^{(0)}_{x_1}\,{u^{(0)}_{x_1}}^* \\
    & = \frac {1}{p(x_1)}\, u^{(0)}_{x_1}
    \sum_{x_3,\ldots,x_n} {t^{(0)}_{x_1}}^*\,{t^{(x_1)}_{x_2}}^* \cdots
    {t^{(x_1,\ldots,x_{n-1})}_{x_n}}^*t^{(x_1,\ldots,x_{n-1})}_{x_n} \cdots t^{(x_1)}_{x_2} \,
    t^{(0)}_{x_1}\,{u^{(0)}_{x_1}}^* \\
    &= \frac {1}{p(x_1)}\, \sum_{x_3,\ldots,x_n} p(x_1,\ldots,x_n)\, \idty
    = p(x_2|x_1)\, \idty\;,
\end{split}\end{equation}
where $p(x_2|x_1)$ is the probability of observing $x_2$ at the second step conditioned upon the
observation of $x_1$ at the first step. Again we have $t^{(x_1)}_{x_2} = \sqrt{p(x_2|x_1)} \,
u^{(x_1)}_{x_2}$ for some unitary $u^{(x_1)}_{x_2}$ and, repeating the same argument, we
find for every $k$
\begin{equation}\label{f2}
    t^{(x_1,\ldots,x_{k-1})}_{x_k} = \sqrt{p(x_k|x_1,\ldots,x_{k-1})}\,
    u^{(x_1,\ldots,x_{k-1})}_{x_k}\;,
    \qquad u^{(x_1,\ldots,x_{k-1})}_{x_k}\in\mathcal{U}(\HH)\;,
\end{equation}
where $p(x_k|x_1,\ldots,x_{k-1})$ is the probability of observing $x_k$ at the $k^{\mathrm{th}}$ step
conditioned upon the observation of $(x_1,\ldots,x_{k-1})$ during the previous steps. Equation \eqref{f2}
gives the polar decomposition of $t^{(x_1,\ldots,x_{k-1})}_{x_k}$, which in this case is unique.

\medskip

(b) Combining equations \eqref{f1} and \eqref{f2} we get
\begin{equation}\nonumber
    t^{(x_1,\ldots,x_{k-1})}_{x_k} = \sqrt{p(x_k|x_1,\ldots,x_{k-1})}\,
    \e^{\im\theta(x_1,\ldots,x_n)} \, {u^{(x_1,\ldots,x_k)}_{x_{k+1}}}^*\cdots
    {u^{(x_1,\ldots,x_{n-1})}_{x_n}}^*\,{u^{(0)}_{x_1}}^*\cdots
    {u^{(x_1,\ldots,x_{k-2})}_{x_{k-1}}}^*\;,
\end{equation}
where, because of the uniqueness of the polar decomposition of $t^{(x_1,\ldots,x_{k-1})}_{x_k}$,
\begin{equation}\nonumber
    \e^{\im\theta(x_1,\ldots,x_n)} \,
    {u^{(x_1,\ldots,x_k)}_{x_{k+1}}}^* \cdots {u^{(x_1,\ldots,x_{n-1})}_{x_n}}^*
    \, {u^{(0)}_{x_1}}^*\cdots {u^{(x_1,\ldots,x_{k-2})}_{x_{k-1}}}^* =
    u^{(x_1,\ldots,x_{k-1})}_{x_k} \;,
\end{equation}
which therefore
has to be independent of $x_{k+1},\ldots,x_n$. If moreover $(x_1,\ldots,x_{k-1})$ is such that for all
$m>k$ the operators
$t^{(x_1,\ldots,x_{m-1})}_{x_m}$ do not depend on $x_k$, then
the same is true for
%$p(x_m|x_1,\ldots,x_{m-1})$, $u^{(x_1,\ldots,x_{m-1})}_{x_m}$ and also for
the unitary operator ${u^{(x_1,\ldots,x_k)}_{x_{k+1}}}^* \cdots
{u^{(x_1,\ldots,x_{n-1})}_{x_n}}^* \, {u^{(0)}_{x_1}}^*\cdots {u^{(x_1,\ldots,x_{k-2})}_{x_{k-1}}}^*$
and so the Kraus operators $t^{(x_1,\ldots,x_{k-1})}_{x_k}$ are all
proportional to a same unitary operator $u^{(x_1,\ldots,x_{k-1})}$ and $T^{(x_1,\ldots,x_{k-1})}(\rho) =
u^{(x_1,\ldots,x_{k-1})}\,\rho\,{u^{(x_1,\ldots,x_{k-1})}}^*$.
\end{proof}

\bigskip

\par\noindent\textit{Proof of Proposition \ref{divPCQI}.\ }\\
\indent (a) $\Rightarrow$ (b). Let $R^{(x_1,\ldots,x_{k-1})}(\rho) = \sum_{j_k}
r^{(x_1,\ldots,x_{k-1})}_{j_k} \,\rho\, {r^{(x_1,\ldots,x_{k-1})}_{j_k}}^*$,
$\sum_{j_k}{r^{(x_1,\ldots,x_{k-1})}_{j_k}}^*r^{(x_1,\ldots,x_{k-1})}_{j_k} = \idty$.
If $\sum_{\substack{x_1,\ldots,x_n\\j_1,\ldots,j_n}} r^{(x_1,\ldots,x_n)}_{j_n}\,t^{(n)}_{x_n}\cdots
r^{(x_1)}_{j_1}\,t^{(1)}_{x_1} \,\rho\,
{t^{(1)}_{x_1}}^*\,{r^{(x_1)}_{j_1}}^*\cdots{t^{(n)}_{x_n}}^*\,{r^{(x_1,\ldots,x_n)}_{j_n}}^* = \rho$ for
every $\rho$, then this decomposition of the ideal channel is associated to a probability measure
$p(x_1,j_1,\ldots,x_n,j_n)$ and Proposition \ref{divQIrest0} directly implies $t^{(k)}_{x_k} =
\sqrt{p(x_k|x_1,j_1,\ldots,x_{k-1},j_{k-1})}\,u^{(x_1,j_1,\ldots,x_{k-1},j_{k-1})}_{x_k}$, where
$u^{(x_1,j_1,\ldots,x_{k-1},j_{k-1})}_{x_k}$ is unitary
% \in \mathcal{U}(\HH)$
and where
$p(x_k|x_1,j_1,\ldots,x_{k-1},j_{k-1})=p^{(k)}_{x_k}$ and
$u^{(x_1,j_1,\ldots,x_{k-1},j_{k-1})}_{x_k}= u^{(k)}_{x_k}$
because $t^{(k)}_{x_k}$ does not depend on $(x_1,j_1,\ldots,x_{k-1},j_{k-1})$.

\medskip

(b) $\Rightarrow$ (c). If $t^{(k)}_{x_k}= \sqrt{p^{(k)}_{x_k}}\,u^{(k)}_{x_k}$, with
$u^{(k)}_{x_k}\in\mathcal{U}(\HH)$, $p^{(k)}_{x_k}\geq0$, $\sum_{x_k} p^{(k)}_{x_k}=1$, then for every
$\rho$
\begin{equation}\nonumber
    \prob(X_1=x_1,\ldots,X_n=x_n) = \tr t^{(n)}_{x_n} \cdots t^{(1)}_{x_1}\, \rho \,
    {t^{(1)}_{x_1}}^* \cdots {t^{(n)}_{x_n}}^* = \prod_{k=1}^n p^{(k)}_{x_k}\;.
\end{equation}

\medskip

(c) $\Rightarrow$ (a). $\tr{t^{(1)}_{x_1}}^* \cdots {t^{(n)}_{x_n}}^*t^{(n)}_{x_n} \cdots t^{(1)}_{x_1}
\,\rho = \tr t^{(n)}_{x_n} \cdots t^{(1)}_{x_1}\, \rho \, {t^{(1)}_{x_1}}^* \cdots {t^{(n)}_{x_n}}^* =
\prob(X_1=x_1,\ldots,X_n=x_n)$ which does not depend on
$\rho$ and so ${t^{(1)}_{x_1}}^* \cdots {t^{(n)}_{x_n}}^*t^{(n)}_{x_n} \cdots t^{(1)}_{x_1}=
\prob(X_1=x_1,\ldots,X_n=x_n)\,\idty$. Then, with the same argument used in the proof of
property (a) of Proposition \ref{divQIrest0},
\begin{equation}\nonumber
    t^{(k)}_{x_k} = \sqrt{\prob(X_k=x_k|X_1=x_1,\ldots,X_{k-1}=x_{k-1})} \,
    u^{(x_1,\ldots,x_{k-1})}_{x_k}\;,
\end{equation}
where $u^{(x_1,\ldots,x_{k-1})}_{x_k}$ is unitary, and where
$\prob(X_k=x_k|X_1=x_1,\ldots,X_{k-1}=x_{k-1})=\prob(X_k=x_k)$ and
$u^{(x_1,\ldots,x_{k-1})}_{x_k}=u^{(k)}_{x_k}$
because $t^{(k)}_{x_k}$ is independ of $(x_1,\ldots,x_{k-1})$. Then the channels
$R^{(x_1,\ldots,x_k)}(\rho)={u^{(k)}_{x_k}}^*\,\rho\,u^{(k)}_{x_k}$ perfectly restore quantum
information.

\medskip

When (a), (b) and (c) hold, all the restoring channels $R$ in (a) satisfy \eqref{divpF} by Proposition
\ref{divQIrest0}, and of course for every choice of $v^{(x_1, \ldots, x_k)}$, $k=1,\ldots,k-1$, the
channel $R^{(x_1,\ldots,x_n)}$ defined by \eqref{divLpF} can perfectly restore quantum information.
Finally equation \eqref{divopl} directly follows from condition (b) and equation \eqref{divpF}.
\hfill $\square$ \vspace{1em}

\section{Optimal recovery of Quantum Information}\label{OptRec}

When perfect correction of quantum information is not possible,
we would like the restoring channels $R$ which bring the corrected channel $\Tcorr$
as close as possible to $\id$, in some sense. We look for channels $R$ which maximize the
\textit{channel fidelity} of the corrected channel, $\fid(\Tcorr)$.
For a channel $T:\BB(\HH)\to\BB(\HH)$,
$T(\rho)=\sum_x t_x\,\rho\,t_x^*$, $\dim\HH=N$, denoted by $\Omega$ a maximally entangled unit vector
in $\HH\otimes\HH$, the channel fidelity
\begin{equation}\nonumber
    \fid(T) = \langle\Omega,T\otimes\id(\ketbra{\Omega}{\Omega})\,\Omega\rangle
    = \frac{1}{N^2} \sum_x |\tr t_x|^2
\end{equation}
measures how well $T$ preserves quantum information, reaching 1 if and only if $T=\id$.

\subsection{Single-step channel}

Let us recall what was found in \cite{GW} for the single-step case.
Given $T$, chosen a Kraus decomposition \eqref{Tdec} and the
restoring channels $R^{(x)}$,
\begin{equation}\nonumber
    R^{(x)}(\rho)=\sum_j r^{(x)}_j \,\rho\, {r^{(x)}_j}^*\;, \qquad
    \sum_j {r^{(x)}_j}^*r^{(x)}_j=\idty\;,
\end{equation}
we are interested in
\begin{equation}\label{FidVal}
    \fid(\Tcorr) = \frac{1}{N^2} \sum_{x,j} |\tr r^{(x)}_j\,t_x|^2\;,
\end{equation}
and we want to maximize it with respect to all possible families $\{R^{(x)}\}$.
\begin{Pro}\label{ORQI} Let $T:\BB(\HH)\to\BB(\HH)$ be a channel over a finite dimensional Hilbert space
$\HH$. Fixed a Kraus decomposition
$T(\rho) = \sum_x t_x\,\rho\,t_x^*$, for every family of channels $R^{(x)}:\BB(\HH)\to\BB(\HH)$ let
$\Tcorr$ be the corresponding overall corrected channel \eqref{Tcorr}. Then
\begin{equation}\label{FidMaj}
    \fid(\Tcorr) \leq \frac{1}{N^2} \sum_x \Big(\tr |t_x| \Big)^2\;.
\end{equation}
Moreover equality holds if and only if
\begin{equation}\label{OTcorr}
    \Tcorr(\rho) = \sum_x |t_x| \,\rho\, |t_x| \;,
\end{equation}
and it can always be attained choosing
\begin{equation}\label{OF}
    R^{(x)}(\rho) = u_x^*\,\rho\,u_x\;,
\end{equation}
where $u_x$ is a unitary operator in the polar decomposition $t_x=u_x\,|t_x|$.
\end{Pro}
\begin{proof} Inequality \eqref{FidMaj} follows from \eqref{FidVal} because the
Cauchy-Schwarz inequality for the Hilbert-Schmidt inner product gives
\begin{equation}\label{CS}
    \sum_j |\tr r^{(x)}_j\,t_x|^2
    = \sum_j\Big|\tr \big(r^{(x)}_j\,u_x|t_x|^{1/2} |t_x|^{1/2}\big) \Big|^2
    \leq \sum_j \tr \big(|t_x|^{1/2}u_x^*{r^{(x)}_j}^*r^{(x)}_j u_x|t_x|^{1/2}\big) \, \tr|t_x|
    = \Big( \tr|t_x|\Big)^2 \;.
\end{equation}

Equality holds in \eqref{FidMaj} if and only if equality holds in \eqref{CS} for every $x$,
which occurs if and only if $r^{(x)}_j\,u_x|t_x|^{1/2} =
\lambda_{xj}|t_x|^{1/2}$, $\lambda_{xj}\in\mathbb{C}$, for
every $x$ and $j$. Then $r^{(x)}_j\,t_x = \lambda_{xj}|t_x|$ and \eqref{OTcorr} holds.
The opposite implication is obvious.

Finally, if $R^{(x)}$ are chosen according to \eqref{OF}, then $\Tcorr$ is given by \eqref{OTcorr}
and equality holds in \eqref{FidMaj}.
\end{proof}

The structure of the optimal restoring channels \eqref{OF} obtained by
maximizing $\fid(\Tcorr)$, is just what one could expect. When a measurement on the environment has given
the result $x$, we deal with a subensemble of systems undergone the state transformation
$\rho\mapsto t_x\,\rho\,t_x^* = u_x |t_x|\,\rho\,|t_x|u_x^*$, which can be
seen as a composition of $\rho\mapsto |t_x|\,\rho\,|t_x|$ followed by $\rho\mapsto
u_x\,\rho\,u_x^*$. Unless we are in the trivial case $|t_x|\propto\idty$, only the second
transformation is physically reversible, and this is just what the channels \eqref{OF} do.
Therefore $\fid(\Tcorr)$ can be maximized considering only unitary feedback, even if sometimes non
unitary feedback could work as well (for example, every time the polar decomposition of $t_x$ is not
unique).

\subsection{Multi-step channel}

Given a multi-step channel $T$, decomposed according to \eqref{divTdec}, we allow corrections also
before the evolution stops and so we want to maximize the channel fidelity of $\Tcorr$ over
all possible families $R^{(x_1)}$, $R^{(x_1,x_2)}$, \ldots, $R^{(x_1,\ldots,x_n)}$ in \eqref{divTcorr}.
We are interested in the optimal feedback, of course,
and we want to compare the maximum attainable fidelity with:
\begin{itemize}
\item the maximum fidelity obtainable only with a unique correction at the end, which is, by
Proposition \ref{ORQI},
\begin{equation}\label{fid'}
    \fid(\Tcorr') = \frac{1}{N^2} \sum_{x_1,\ldots,x_n} \Big(\tr \big|t^{(n)}_{x_n} \cdots
    t^{(1)}_{x_1}\big|\Big)^2\;,
\end{equation}
where $\Tcorr'$ denotes the corrected channel
\begin{equation}\nonumber
    \Tcorr'(\rho) = \big|t^{(n)}_{x_n} \cdots t^{(1)}_{x_1}\big| \,\rho\,
    \big|t^{(n)}_{x_n} \cdots t^{(1)}_{x_1}\big|\;;
\end{equation}
\item the fidelity obtainable applying step by step the feedback suggested by Proposition \ref{ORQI},
chosen on the basis of the whole avaible information; the corrected evolution associated to the
observation of $(x_1,\ldots,x_n)$ would be
\begin{equation}\label{sbsF}\begin{split}
    \rho &\mapsto t^{(1)}_{x_1}\, \rho \, {t^{(1)}_{x_1}}^*
    \mapsto |t^{(1)}_{x_1}|\, \rho \, |{t^{(1)}_{x_1}}|
    \mapsto t^{(2)}_{x_2}\,|t^{(1)}_{x_1}|\, \rho \, |{t^{(1)}_{x_1}}|\,{t^{(2)}_{x_2}}^* \mapsto
    \big|t^{(2)}_{x_2}\,|t^{(1)}_{x_1}|\big| \,\rho\, \big|t^{(2)}_{x_2}\,|t^{(1)}_{x_1}|\big|
    \mapsto \cdots \\
    &\mapsto \Big|t^{(n)}_{x_n}\cdots\big|t^{(2)}_{x_2}|t^{(1)}_{x_1}|\big|\Big| \,\rho\,
    \Big|t^{(n)}_{x_n}\cdots\big|t^{(2)}_{x_2}|t^{(1)}_{x_1}|\big|\Big| := \Tcorr''(\rho)\;,
\end{split}\end{equation}
with fidelity
\begin{equation}\label{fid''}
    \fid(\Tcorr'') = \frac{1}{N^2} \sum_{x_1,\ldots,x_n}
    \Big(\tr \Big|t^{(n)}_{x_n} \cdots \big|t^{(2)}_{x_2} |t^{(1)}_{x_1}| \big| \Big| \Big)^2\;.
\end{equation}
\end{itemize}
Quite surprisingly, these two strategies are equivalent, and optimal, in the qubit case ($\dim\HH=2$),
but in the general case ($\dim\HH\geq3$) there is no relationship between them, and neither of them
gives the optimal correction.

\subsubsection{Qubit multi-step channels}

\begin{Pro}\label{divQbORQI} Let $\dim\HH=2$ and $T=T^{(n)}\circ\cdots\circ T^{(1)}$ be a multi-step
channel, with every $T^{(k)}:\BB(\HH)\to\BB(\HH)$. Fixed the Kraus decompositions
$T^{(k)}(\rho)=\sum_{x_k} t^{(k)}_{x_k}\,\rho\,{t^{(k)}_{x_k}}^*$,
for every family of channels $R^{(x_1)}$, $R^{(x_1,x_2)}$, \ldots, $R^{(x_1,\ldots,x_n)}$, let
$\Tcorr$ be the corresponding overall corrected channel \eqref{divTcorr}. Then
\begin{equation}\label{divQbFidMaj}
    \fid(\Tcorr) \leq \frac{1}{4} \sum_{x_1,\ldots,x_n}
    \Big(\tr \big|t^{(n)}_{x_n} \cdots t^{(1)}_{x_1}\big| \Big)^2\;.
\end{equation}
Moreover equality can always be attained choosing
\begin{equation}\label{divQbOF}
    R^{(x_1, \ldots, x_k)}(\rho)=v^{(x_1, \ldots, x_k)}\,\rho\,{v^{(x_1, \ldots, x_k)}}^*\;, \qquad
    v^{(x_1, \ldots, x_k)} \in \mathcal{U}(\HH)\;, \qquad \forall\;k=1,\ldots,n\;,
\end{equation}
with $v^{(x_1, \ldots, x_n)}={w^{(x_1, \ldots, x_n)}}^*$, where $w^{(x_1, \ldots, x_n)}$ is a unitary
operator in the polar decomposition $t^{(n)}_{x_n}\,v^{(x_1, \ldots, x_{n-1})} \cdots
v^{(x_1)}\,t^{(1)}_{x_1} = w^{(x_1, \ldots, x_n)}\,\big|t^{(n)}_{x_n}\,v^{(x_1, \ldots, x_{n-1})} \cdots
v^{(x_1)}\,t^{(1)}_{x_1}\big|$.
\end{Pro}
\begin{proof} The key property in a 2-dimensional Hilbert space is that for every operator $A$
\begin{equation}\nonumber
    (\tr|A|)^2 = \tr(|A|^2) + 2\det|A|\;, \qquad \mbox{ where } \det|A|=|\det A| \leq \frac{1}{2}
    \tr(|A|^2)\;.
\end{equation}
Then, for every family of channels $R^{(x_1,\ldots,x_{k-1})}(\rho) = \sum_{j_k}
r^{(x_1,\ldots,x_{k-1})}_{j_k} \,\rho\, {r^{(x_1,\ldots,x_{k-1})}_{j_k}}^*$,
$\sum_{j_k}{r^{(x_1,\ldots,x_{k-1})}_{j_k}}^*r^{(x_1,\ldots,x_{k-1})}_{j_k} = \idty$,
\begin{equation}\nonumber\begin{split}
    \fid(\Tcorr) &= \frac{1}{4} \sum_{\substack{x_1,\ldots,x_n\\j_1,\ldots,j_n}}
    \big|\tr r^{(x_1,\ldots,x_n)}_{j_n}\,t^{(n)}_{x_n}\cdots r^{(x_1)}_{j_1}\,t^{(1)}_{x_1}\big|^2
    \leq \frac{1}{4} \sum_{\substack{x_1,\ldots,x_n\\j_1,\ldots,j_{n-1}}}
    \Big(\tr \big|t^{(n)}_{x_n}\,r^{(x_1,\ldots,x_{n-1})}_{j_{n-1}}\cdots
    r^{(x_1)}_{j_1}\,t^{(1)}_{x_1}\big| \Big)^2
    \\
    &=\frac{1}{2} + \frac{1}{2} \sum_{\substack{x_1,\ldots,x_n\\j_1,\ldots,j_{n-1}}}
    \big|\det t^{(n)}_{x_n}\cdots t^{(1)}_{x_1} \big| \cdot \big|\det
    r^{(x_1,\ldots,x_{n-1})}_{j_{n-1}} \big| \cdot
    \big|\det r^{(x_1)}_{j_1}\big|
    \leq \frac{1}{2} + \frac{1}{2} \sum_{x_1,\ldots,x_n}
    \big|\det t^{(n)}_{x_n}\cdots t^{(1)}_{x_1} \big| \\
    &= \frac{1}{4} \sum_{x_1,\ldots,x_n} \Big(\tr \big|t^{(n)}_{x_n} \cdots t^{(1)}_{x_1}\big|
    \Big)^2\;.
\end{split}\end{equation}
Analogously for every family of unitary channels $R^{(x_1,\ldots,x_k)}(\rho) =
v^{(x_1, \ldots, x_k)}\,\rho\,{v^{(x_1, \ldots, x_k)}}^*$, with $v^{(x_1, \ldots, x_n)} =
{w^{(x_1,\ldots,x_n)}}^*$,
\begin{equation}\nonumber\begin{split}
    \fid(\Tcorr) &= \frac{1}{4} \sum_{x_1,\ldots,x_n}
    \Big(\tr \big|t^{(n)}_{x_n}\,v^{(x_1,\ldots,x_{n-1})}\cdots v^{(x_1)}\,t^{(1)}_{x_1}\big|\Big)^2
    =\frac{1}{2} + \frac{1}{2} \sum_{x_1,\ldots,x_n}
    \big|\det t^{(n)}_{x_n}\,v^{(x_1,\ldots,x_{n-1})}\cdots v^{(x_1)}\, t^{(1)}_{x_1} \big| \\
    &=\frac{1}{2} + \frac{1}{2} \sum_{x_1,\ldots,x_n}
    \big|\det t^{(n)}_{x_n}\cdots t^{(1)}_{x_1} \big|
    = \frac{1}{4} \sum_{x_1,\ldots,x_n} \Big(\tr \big|t^{(n)}_{x_n} \cdots t^{(1)}_{x_1}\big|\Big)^2
    \;.
\end{split}\end{equation}
\end{proof}

\subsubsection{Higher dimensional two-step channels}
\label{optimult}

When $\dim\HH\geq3$ there is no general relationship between the
two fidelities \eqref{fid'} and \eqref{fid''}. Consider on
$\mathbb{C}^3$ the multi-step channels $T=T^{(2)}\circ T^{(1)}$
and $S=S^{(2)}\circ S^{(1)}$, where
\begin{gather}
    T^{(1)}(\rho) = T^{(2)}(\rho) = \sum_{x=1}^2 t_x \, \rho \, t_x^*\;, \qquad \qquad
    S^{(1)}(\rho) = S^{(2)}(\rho) = \sum_{x=1}^2 s_x \, \rho \, s_x^*\;, \nonumber\\
    |t_1|^2=|s_1|^2=\begin{pmatrix}1/6&0&0\\0&1/3&0\\0&0&1/2\end{pmatrix}\;, \qquad \qquad
    |t_2|^2=|s_2|^2=\begin{pmatrix}5/6&0&0\\0&2/3&0\\0&0&1/2\end{pmatrix}\;,\nonumber\\
    t_1= u\,|t_1|\;, \qquad t_2= |t_2|\;, \qquad\qquad
    s_1= v\,|s_1|\;, \qquad s_1= u\,|s_1|\;,\nonumber
\end{gather}
  where
\begin{equation}
    u=\begin{pmatrix}0&1&0\\1&0&0\\0&0&1\end{pmatrix} \;, \qquad\mbox{and} \qquad
    v= \frac{1}{\sqrt{3}} \begin{pmatrix}\e^{2\im\pi/3}&\e^{-2\im\pi/3}&1\\
    \e^{-2\im\pi/3}&\e^{2\im\pi/3}&1\\1&1&1\end{pmatrix} \;. \nonumber
\end{equation}
Then since the absolute values of the Kraus operators coincide,
the single-step corrected versions of $T$ and $S$ are the same.
Moreover, since these absolute values commute, the iterated
absolute values factorize, i.e.,
\begin{equation}
    \Big|t^{(n)}_{x_n}\cdots\big|t^{(2)}_{x_2}\,|t^{(1)}_{x_1}|\big|\big|
    =|t^{(n)}_{x_n}|\cdots|t^{(2)}_{x_2}|\cdot|t^{(1)}_{x_1}|
\end{equation}
Hence the greedy strategy of correcting for maximal fidelity at
every step, using all available information, gives the same result
for these channels as the product of the single-step corrected
channels. In particular, $S$ and $T$ are equivalent in this
respect, and $\fid(\Tcorr'')=\fid(\Scorr'')$ (see last line in the
table below).

On the other hand, we can leave the system undisturbed, and only
make an optimal correction at the end, again using all available
information. Finally, we may optimize fidelity over all correction
schemes. Numerically this achieved by an iteration, developed in
\cite{RW}, which efficiently determines the maximum of a general
positive linear functional on channels. In the case at hand this
is the overall channel fidelity as a function of the correcting
channels $R$.  Since we can correct the unitaries $u$ and $v$
immediately in every step, it is clear that the channels $S$ and
$T$ once again give the same corrected fidelities. The results are
summarized in the following table:

\vskip12pt

\begin{center}
\begin{tabular}{|l|c|c|}
 Channel fidelities& \qquad$T$\qquad{}     &\qquad$S$\qquad{} \\\hline
 optimal correction             &0.9584 & 0.9584   \\
 final correction only          &0.9556 & 0.9576   \\
                                & $<$   &  $>$     \\
 optimal correction at each step&0.9570 & 0.9570    \\
\end{tabular}
\end{center}

Hence, in contrast to the case of perfect corrigibility, it may
help to perform corrections on the way, rather than a single
correction at the end (column $T$). Perhaps surprisingly, the
optimal correction at the intermediate steps requires some
foresight, and depends on what is to follow: otherwise the
strategy of correcting for highest fidelity using all available
information, which is clearly optimal in the last step, would also
have to be optimal in the intermediate steps (compare first and
last line). Even leaving out the intermediate correction
altogether may be better (column $S$).

\subsubsection{Higher dimensional many-step channels}

We noted that in the previous example the correction scheme for
$\Tcorr''$ is equivalent to multiplying the one-step corrected
channels. That is, whether this correction step may use all
available information or just the information from the last step,
leads to the same corrected channel. We showed this by noting that
the absolute values of all Kraus operators involved commute. This
will be true in any dimension if there are only two Kraus
operators, as in the example. Here we just note that this equality
is not a general fact. In fact using random channels one quickly
finds examples (3 Hilbert space dimensions, 3 Kraus operators, and
4 time steps), in which using all information or just information
from the last step give different results. Surprisingly, once
again the comparison may be either way: sometimes using only
one-step information is {\it better} than using all information.
Of course, this is only possible because the greedy strategy is
not optimal in the first place.

\section{Outlook}
We would like to spend some words about the quantum error-correction in \textit{continuous} time.
Our aim would be to generalize this correction scheme for quantum information corrupted by a
noisy Markovian evolution assigned by a Quantum Dynamical Semigroup.
The present analysis for a multi-step channel $T= \big(T^{(1)}\big)^n$
would provide the intervalwise treatment, but it is not easy to take
the limit to continuous time. The well established theory of continuous measurements gives the
possible decompositions (unravelings) of the channel according to the observed trajectoies,
but a continuous time formulation of general non-Hamiltonian feedback is missing.
There is no general description of the
evolution conditioned on the observed process if a feedback depending on the whole measurement record
is added. And also if the feedback is supposed to depend only on the present observation, the measured
current, still there is no general description for non-Hamiltonian feedback. A theory exists only for
Hamiltonian feedback simply proportional to the measured currents \cite{Wiseman}.
Anyway a preliminary study of perfect correction in this framework leads to the same conclusions of
Proposition \ref{divPCQI}: perfect correction is possible if and only if the measured process is
uninformative and in this case a unique correction at the end suffices. It is also possible to
characterize the Linblad generator of semigroups allowing such perfect correction and, as pointed out
by Luc Bouten, these semigroups are just the ones admitting an essentially commutative dilation
\cite{KM}.

\bigskip

\noindent\textbf{Acknowledgments.} M.\ G.\ gratefully acknowledges
support from the Alexander von Humboldt Stiftung. We thank Michael
Reimpell for computing the optimal corrections in
Section~\ref{optimult}.

%%%%%%%%%%%%%%%%%%%%%%%%%%%%%%%%%%%%%%%%%%%%%%%%%%%%%%%%%%%%%%%%%%%%%%%%%%%%%%%

\end{document}